\documentclass[floatfix,pra,aps,a4paper,twocolumn,showpacs]{revtex4}

 \usepackage{bm}
 \usepackage{amsmath}
 \usepackage{amssymb}
 \usepackage{latexsym}
 \usepackage{amsfonts}
 \usepackage{epsfig}

 \newcommand{\expval}[1]{\left< #1 \right>}

\newcommand{\ket}[1]{\left|#1\right>}
\newcommand{\bra}[1]{\left<#1\right|}
 
 \newcommand{\nn}{\nonumber\\}
 
 \newcommand{\f}[1]{\mbox{\boldmath$#1$}}

 \newcommand{\bea}{\begin{eqnarray}}
 \newcommand{\ea}{\end{eqnarray}}
 \newcommand{\eea}{\end{eqnarray}}
 
 \newcommand{\trace}[1]{{\rm Tr}\left\{ #1 \right\}}
 \newcommand{\abs}[1]{{\left| #1 \right|}}

 \newcommand{\HS}{H_{\rm S}}
 \newcommand{\HB}{H_{\rm B}}
 \newcommand{\HI}{H_{\rm SB}}

 \newcommand{\sgn}[1]{{\rm sgn}\left( #1 \right)}
 
 \newcommand{\ii}{{\rm i}}

\begin{document}

\title{Fighting Decoherence by Feedback-controlled Dissipation}

\author{Gernot Schaller}
\email{gernot.schaller@tu-berlin.de}


\affiliation{Institut f\"ur Theoretische Physik, Technische Universit\"at Berlin, Hardenbergstr. 36, 10623 Berlin, Germany}

\begin{abstract}
Repeated closed-loop control operations acting as piecewise-constant Liouville superoperators conditioned on the outcomes of
regularly performed measurements may effectively be described by a fixed-point iteration for the density matrix.
Even when all Liouville superoperators point to the completely mixed state, feedback of the measurement result
may lead to a pure state, which can be interpreted as selective dampening of undesired states.
Using a microscopic model, we exemplify this for a single qubit, which can be purified in an arbitrary single-qubit state
by tuning the measurement direction and two qubits that may be purified towards a Bell state by applying a special continuous 
two-local measurement.
The method does not require precise knowledge of decoherence channels and works for large reservoir temperatures 
provided measurement, processing, and control can be implemented in a continuous fashion.
\end{abstract}

\maketitle

In quantum systems, one typically aims at avoiding decoherence that is often seen as arch-enemy of quantum computation~\cite{nielsen2000}.
Simply performing the computation fast enough will fail for most applications as a size-scalable quantum system with large 
coherence times~\cite{divincenzo2000} is yet to be found.
Therefore, advanced schemes have been proposed to reduce or inhibit decoherence in quantum computers such as e.g.
quantum error correction~\cite{shor1995a}, the use of decoherence-free subspaces~\cite{zanardi1997a}, or 
open-loop control~\cite{viola1999a,platzer2010a,paz_silva2012a,hwang2012a}.
These schemes require quite sophisticated techniques which themselves may not be entirely robust against decoherence and/or control errors.

The idea to revert the perspective by using decoherence constructively is not new~\cite{rubin1982a,poyatos1996a,beige2000a,lloyd2001a} but has 
recently again gained a lot of attention~\cite{diehl2008a,verstraete2009b,vollbrecht2011a,ticozzi2012a,koga2012a}:
The simplest example is to relax into the (possibly entangled) ground state of a defined system Hamiltonian, which only requires weak unspecific couplings 
to a low-temperature reservoir with a sufficiently large energy gap above the ground $\Delta E \gg k_{\rm B} T$.
The limitation to the ground state can be overcome by using multiple reservoirs at different thermal equilibria and different chemical potentials.
Then however, the general dissipative engineering paradigm requires to design the interactions such that 
the system is driven to the desired target state.
For a general pure state to be stabilized, this problem is hard to solve, the solution is 
specific to the target state and may prove difficult to implement experimentally.

Since the success of the centrifugal governor used by James Watt to regulate the speed of steam engines, 
{\em feedback} (closed-loop) control has found wide-spread application with todays standard 
examples including e.g. thermostats and automatic speed control in cars.
Nowadays, the built-in mechanical self-regulation of the centrifugal governor is often replaced by electronic signal processing,
which enables one to change the feedback protocol easily.
The implementation of quantum feedback control requires the inclusion of the quantum-mechanical
measurement process and to make the quantum-mechanical evolution after each measurement subject to control.
In the conventional scheme~\cite{wiseman1993a,wiseman1994a}, quantum jumps detected during the time-interval $\Delta t$ 
are fed back as an instantaneous $\delta$-pulse into the system Hamiltonian.
In the continuum measurement limit $\Delta t \to 0$, an effective master equation emerges, where the control operation appears as an 
instantaneous unitary rotation of the density matrix~\cite{wiseman2010,kiesslich2011a,poeltl2011a}.

In contrast, here we consider a general scheme of periodic measurements at intervals $\Delta t$ 
and discuss in detail the opposite case where the collapse due to projective measurements is assumed to be much faster
than $\Delta t$ throughout.
The scheme allows for weak measurements and in principle arbitrary control pulse sequences, but we
will for simplicity of interpretation specialize on a piecewise-constant time-dependence of the control parameters.
In this scenario, even for $\Delta t \to 0$ a master equation description does not typically emerge.
The purpose of this paper is to show that, provided with dissipation that drags the system only 
to the completely mixed state but allows for strength-tunable system-reservoir interactions, feedback of 
information obtained by measurements can be used to achieve purification and thereby puts 
even ''clueless'' dissipation to a productive role.

\section{Propagation under Feedback}

A general quantum measurement is described by a set of measurement operators $\{M_m\}$ satisfying the
completeness relation $\sum_m M_m^\dagger M_m = \f{1}$~\cite{nielsen2000}.
Orthogonal projection operators $M_n M_m = \delta_{nm} M_n$ are a well-known standard case.
Under measurement outcome $m$, the density matrix becomes (we assume an instantaneous collapse
of the wave function)
\bea
\rho \stackrel{m}{\to} \frac{M_m \rho M_m^\dagger}{P(m)}\,,
\eea
and the outcome probability is given by $P(m)=\trace{M_m^\dagger M_m \rho}$.
Arranging the $N^2$ elements of the density matrix in a vector, this becomes
\bea
\rho \stackrel{m}{\to} \frac{1}{P(m)} {\cal M}_m \rho\,,
\eea
where ${\cal M}_m$ is a superoperator having an $N^2 \times N^2$ matrix representation.
In the following, we will use calligraphic symbols to denote superoperators.
In principle, the measurement operators may depend also on the time interval $\Delta t$ between two measurements, see
appendix~\ref{ASjumpmeasurement}.

The most general evolution of the density matrix (including measurements, unitary and dissipative evolution) is governed by a trace- and positivity preserving map
\bea
\rho(t+\Delta t) &=& \sum_\alpha B_\alpha(\Delta t) \rho(t) B_\alpha^\dagger(\Delta t)\nn
&=& \sum_\alpha {\cal B}_\alpha(\Delta t) \rho(t) \hat{=} {\cal B}(\Delta t) \rho(t)
\eea
with Kraus operators obeying $\sum_\alpha B_\alpha^\dagger(\Delta t) B_\alpha(\Delta t) = \f{1}$ and the 
corresponding superoperator ${\cal B}(\Delta t)$.
Formally, feedback can be conveniently described by making the general evolution 
between two measurements dependent on the outcome of the previous measurement.
For example, given density matrix $\rho(t)$ right before one measurement, the density matrix conditioned on outcome $m$ at the first measurement 
becomes right before the next measurement
\bea
\rho(t+\Delta t) &\stackrel{m}{\to}& \sum_\alpha B_{\alpha,m}(\Delta t) \frac{M_m \rho(t) M_m^\dagger}{P(m)} B_{\alpha,m}^\dagger(\Delta t)\nn
&\hat{=}& \frac{ {\cal B}_{m}(\Delta t) {\cal M}_m \rho(t)}{P(m)}\,.
\eea
For an arbitrarily chosen observable, we perform a weighted average for its expectation value at time $t+\Delta t$ over all measurement outcomes
and conditioned evolutions
\bea
\expval{\bar{A}_{t+\Delta t}}= \trace{{\cal A} \left[\sum_m {\cal B}_{m}(\Delta t) {\cal M}_m\right]\rho(t)}\,,
\eea
where we have used superoperators for the ease of notation with ${\cal A}\rho \hat{=} A \rho$ and the trace is generally mapped to
multiplication with a row vector containing $N$ entries of value $1$ at positions containing populations and $N^2-N$ entries of value 
$0$ elsewhere.
Demanding \mbox{$\expval{\bar{A}_{t+\Delta t}}\stackrel{!}{=}\trace{{\cal A} \rho(t+\Delta t)}$} defines an effective propagator for repeated measurements and feedback
\bea\label{Epropeffgeneral}
{\cal P}_{\rm eff}(\Delta t) = \sum_m {\cal B}_{m}(\Delta t) {\cal M}_m\,.
\eea
The above propagator may prove useful once microscopic parameters are linked to the superoperators ${\cal B}_{m}(\Delta t)$, which
is however in principle possible also for non-Markovian systems~\cite{lidar2001a,schaller2008a,schaller2009a}.
In appendix~\ref{ASjumpmeasurement} we show that it yields an effective master equation as in the conventional Wiseman-Milburn scheme 
for non-projective quantum jump detection, unitary control and in the continuum limit $\Delta t\to 0$.

In the following however, we will consider a Lindblad evolution with a single constant Lindblad superoperator ${\cal L}_m$
(which may however also include Hamiltonian control~\cite{ticozzi2010a})
between the measurements ${\cal B}_{m}(\Delta t) = e^{{\cal L}_m \Delta t}$ and
time-independent measurement superoperators
\bea\label{Epropeff}
{\cal P}_{\rm eff}(\Delta t) = \sum_m e^{{\cal L}_m \Delta t} {\cal M}_m\,.
\eea
Note that in contrast to the Wiseman-Milburn scheme (see also appendix~\ref{ASjumpmeasurement}), 
the time-independence of the measurement superoperators requires the measurement to be much faster than $\Delta t$ throughout.
The conditioning of the ${\cal L}_m$ on the previous measurement result $m$ defines the feedback protocol.
Microscopically, this dependence on the measurement result can be implemented by triggering switches of parameters in the Hamiltonian.
We explicitly include the possibility that the stationary state  (defined by ${\cal L}_m \bar{\rho}=0$) of each Liouvillian may be the
completely mixed state with $\trace{\bar{\rho}^2}=1/N$.
In the following, we will discuss the implications of this effective propagator for the control of qubits.

Typically, decoherence in an open quantum system is modeled by the decay of off-diagonal matrix elements of the density matrix (coherences).
In an appropriate basis, this is formally represented by a block form of the Liouvillians ${\cal L}_m$, which leads to a decoupling of coherences and diagonal entries (populations).
A prototypical example for this is the Born-Markov-secular approximation master equation in the case of a non-degenerate system Hamiltonian~\cite{breuer2002}, 
where the block structure becomes explicit in the system energy eigenbasis.
However, for quantum measurements the measurement superoperators ${\cal M}_m$ will not generally have the same block structure as the Liouvillian.
It may therefore be expected that the effective propagator in Eq.~(\ref{Epropeff}) couples coherences and populations and thereby may lead to nontrivial effects.

In spite of the completeness relation in the original Hilbert space, the measurement superoperators do not sum up to the identity $\sum_m {\cal M}_m \neq \f{1}$.
This implies that even in the continuum limit, where measurement, processing, and control are repeated at infinitesimally small time intervals $\Delta t\to 0$, the iteration equation
$\rho(t+\Delta t) = {\cal P}_{\rm eff}(\Delta t) \rho(t)$ with (\ref{Epropeff}) does not converge to an effective master equation $\dot{\rho} \neq {\cal L}_{\rm eff} \rho$.
A prominent example for this is the quantum Zeno effect: For a closed system without measurements and without feedback, the action of the Liouvillian without measurement 
simply becomes $\rho(t+\Delta t) = e^{{\cal L}_0 \Delta t} \rho(t) \hat{=} e^{-\ii H \Delta t} \rho e^{+\ii H \Delta t}$, which just encodes the usual unitary evolution.
With measurements and without feedback, the effective propagator in Eq.~(\ref{Epropeff}) leads to
$\rho(t+\Delta t) = e^{{\cal L}_0 \Delta t} \left[\sum_m {\cal M}_m\right]\rho(t)\hat{=} e^{-\ii H \Delta t} \left[\sum_m M_m \rho(t) M_m^\dagger\right] e^{+\ii H \Delta t}$, 
where the measurement (super-) operators may account for the freezing of the observed quantum state when $\Delta t \to 0$, as will be discussed in the next section in greater detail.

A continuous description by means of a differential equation for expectation values may however still be possible when the expectation values are consistent with the 
measurement operators, as will be shown below.

\section{Dissipative Purification of a Qubit}
As a first example we consider a single qubit parameterized by Pauli matrices.
At time intervals $\Delta t$, we perform strong projective measurements of $\sigma^x$ described by the
measurement operators $M_\pm = \frac{1}{2}\left[\f{1}\pm\sigma^x\right]$ with superoperator equivalents
acting on the vector $\left(\rho_{00},\rho_{11},\rho_{01},\rho_{10}\right)^{\rm T}$ having the representation
\bea
{\cal M}_\pm = \frac{1}{4}\left(\begin{array}{cccc}
1 & 1 & \pm1 & \pm1\\
1 & 1 & \pm1 & \pm1\\
\pm1 & \pm1 & 1 & 1\\
\pm1 & \pm1 & 1 & 1
\end{array}\right)\,.
\eea
The feedback protocol is defined by applying an outcome-dependent Liouvillian derived microscopically from the system (S), bath (B) and 
interaction (SB) Hamiltonians
\bea\label{Ehamqubit}
\HS^{(\pm)} &=& \frac{\Omega}{2}\sigma^z\,,\qquad \HB^{(\pm)} = \sum_k \omega_k b_k^\dagger b_k\,,\nn
\HI^{(\pm)} &=& \lambda_\pm \left(\f{n}_\pm \cdot \f{\sigma}\right) \otimes \sum_k \left(h_k b_k + h_k^* b_k^\dagger\right)\,,
\eea
where $\lambda_\pm \ge 0$ and the unit vectors $\f{n}_\pm=\left(\sin(\theta_\pm) \cos(\phi_\pm), \sin(\theta_\pm) \sin(\phi_\pm),\cos(\theta_\pm)\right)$ 
characterize strength and direction (e.g., purely dephasing for $\f{n}=\f{e}_z$)
of the dissipation, respectively.
The $b_k$ are bosonic annihilation operators of a bath assumed at thermal equilibrium throughout.
The case $\lambda_\pm=\lambda$ and $\f{n}_\pm = \f{n}$ corresponds to repeated measurements without feedback applied to an open quantum system.
Applying the Born, Markov and secular approximations, the corresponding conditioned Liouville superoperators ${\cal L}_\pm$ do not
depend on $\phi_\pm$ and both have block structure in the system energy eigenbasis
$\sigma^z\ket{0/1} = (-1)^{(0/1)} \ket{0/1}$, leading to a decoupling of coherences and populations, see appendix~\ref{ASsinglequbit}.

The ability to change the system-reservoir coupling will of course depend on the physical implementation of the qubit.
In electronic setups for example (charge qubits), nearby quantum point contacts may not only function as detector:
Changing the bias voltage for circuits in the vicinity of the charge qubit will induce a changed system-reservoir coupling.
If in contrast the qubit is realized as the lowest two modes of a cavity, tuning the permeability of the cavity walls may yield a
similar effect.

\subsection{Thermal reservoirs}

Let us first consider the simplest case where neither measurement nor control are applied:
The continuous action of either Liouvillian would lead to an exponential decay of coherences
$\abs{\rho_{01}}^2(t) = e^{-\Gamma_\pm t} \abs{\rho_{01}}^2(0)$ with the dephasing rate
\bea\label{Edephrate}
\Gamma_\pm = \lambda_\pm^2\left\{4 \cos^2(\theta_\pm) \gamma_0 + \sin^2(\theta_\pm) \left[\gamma_{+\Omega}+\gamma_{-\Omega}\right]\right\}\,.
\eea
Here,  
$\gamma_\omega \equiv\int \expval{e^{+\ii\HB\tau} B e^{-\ii\HB\tau} B} e^{+\ii\omega\tau}d\tau$ 
is the Fourier transform of the bath correlation function with the coupling operator $B\equiv\sum_k \left(h_k b_k + h_k^* b_k^\dagger\right)$ when
a thermal state at inverse temperature $\beta$ is assumed.
Analytically continuing the spectral coupling density $J(\omega)\equiv2\pi\sum_k \abs{h_k}^2 \delta(\omega-\omega_k)$ to negative $\omega$ via $J(-\omega)\equiv-J(+\omega)$ one 
obtains with the Bose distribution $n_{\rm B}(\omega) = \left[e^{+\beta\omega}-1\right]^{-1}$ the simple expression
$\gamma_\omega= J(\omega) \left[1+n_{\rm B}(\omega)\right]$.
For both Liouvillians, the populations will approach the thermalized stationary state characterized by $\bar{\sigma}^z = (\gamma_{-\Omega}-\gamma_{+\Omega})/(\gamma_{-\Omega}+\gamma_{+\Omega})$.
Due to micro-reversibility, the correlation function obeys the Kubo-Martin-Schwinger~\cite{breuer2002} 
condition $\gamma_{-\Omega} = e^{-\beta\Omega} \gamma_{+\Omega}$.

Therefore, for low temperatures $k_{\rm B} T \ll \Omega$, the qubit will simply decay towards its ground state, which is a trivial example of
dissipation-induced purification in the eigenbasis of the system Hamiltonian.
Note however that the validity of the Markovian approximation is only expected when the coupling strength is smaller than the temperature, 
such that the relaxation speed may be slowed by scaling the coupling strength appropriately.

In contrast, for high temperatures and for an Ohmic spectral density $J(\omega) = 2\alpha \omega e^{-\omega/\omega_{\rm c}}$~\cite{weiss1993}, 
the Fourier transform of the correlation function becomes essentially flat for large cutoff frequencies $\omega_{\rm c}$
with a limiting scaling $\gamma \equiv k_{\rm B} T 2\alpha \approx \gamma_0 \approx \gamma_{\pm\Omega}$.
Here, the Markovian approximation becomes exact as the bath correlation function becomes a Dirac-$\delta$ distribution.
The high-temperature stationary state is then just the completely mixed state with $\expval{\bar{\sigma}^{x/y/z}}=0$ and in essence, 
decoherence occurs more rapidly at higher temperatures.
In this limit also further terms corresponding to the Lamb-shift are negligible.
For simplicity and to stress the drastic merit of feedback control, we will therefore in the following focus on 
this worst-case scenario for decoherence (i.e., the high-temperature and wide-band limit),
where the Fourier transform of the bath correlation function is characterized by the single parameter $\gamma$.

\subsection{Quantum Zeno Limit}

Performing repeated measurements but neither allowing for feedback nor dissipation ($\lambda_\pm=0$), the effective propagator (\ref{Epropeff}) becomes
\bea\label{Epropzeno}
{\cal P}_{\rm Zeno}(\Delta t) = \frac{1}{2}\left(\begin{array}{cccc}
1 & 1 & 0 & 0\\
1 & 1 & 0 & 0\\
0 & 0 & e^{-\ii\Omega\Delta t} & e^{-\ii\Omega\Delta t}\\
0 & 0 & e^{+\ii\Omega\Delta t} & e^{+\ii\Omega\Delta t}
\end{array}\right)\,,
\eea
which simply leads to 
the Quantum Zeno effect~\cite{misra1977a,facchi2002a}.
Repeated application of the above propagator yields
\bea
{\cal P}_{\rm Zeno}^n(\Delta t) = \left(\begin{array}{cc}
\f{1} & \f{0}\\
\f{0} & \cos^{n-1}(\Omega \Delta t) \f{1}
\end{array}\right) {\cal P}_{\rm Zeno}(\Delta t)\,,
\eea
which when $t=n \Delta t$ is kept constant while $\Delta t \to 0$ and $n\to\infty$
reduces further to ${\cal P}_{\rm Zeno}^n(\Delta t) \to {\cal P}_{\rm Zeno}(0)$.
Obviously, this operator preserves the pure eigenstates of the measurement 
operators $\ket{\Psi}=\frac{1}{\sqrt{2}}\left[\ket{0}\pm\ket{1}\right]$,
such that the system when initialized in these states will remain frozen.

\subsection{Zeno and dissipation}

Adding dissipation but without feedback ($\lambda_\pm=\lambda>0$ and $\f{n}_\pm = \f{n}$), the Quantum Zeno effect does not stabilize the coherences: 
According to Eq.~(\ref{Epropeff}),
these decay between the measurements as exemplified by the decaying expectation 
values 
\bea
\expval{\sigma^x_{t+\Delta t}} &=& \cos(\Omega\Delta t) e^{-\gamma \Delta t \lambda^2 [3+\cos(2\theta)]/2} \expval{\sigma^x_t}\,,\nn
\expval{\sigma^y_{t+\Delta t}} &=& \sin(\Omega\Delta t) e^{-\gamma \Delta t \lambda^2 [3+\cos(2\theta)]/2} \expval{\sigma^x_t}\,, 
\eea
whilst $\expval{\sigma^z_{t+\Delta t}}=0$.
Note that the chosen measurement of $\sigma^x$ at time $t$ projects $\expval{\sigma^{y/z}_t}$ to zero, which eventually implies that
only the expectation value of $\expval{\sigma^x_t}$ has a differential equation limit as $\Delta t\to 0$.

\subsection{Feedback}

\begin{figure}[ht]
\includegraphics[width=0.48\textwidth,clip=true]{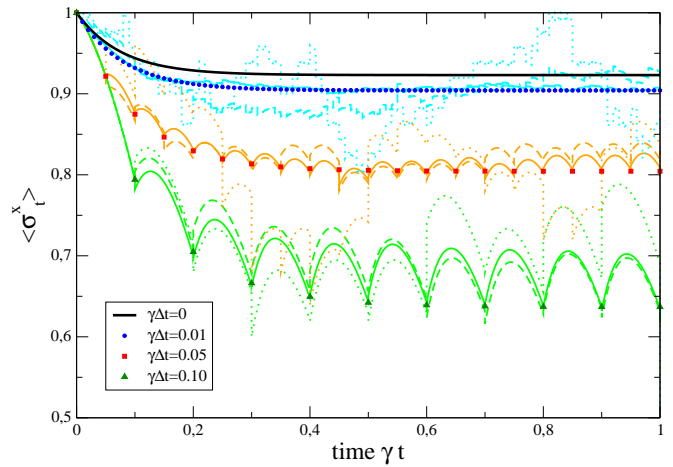}
\caption{\label{Fsqbtraj}(Color Online)
Comparison of the effective evolution of the $\expval{\sigma^x_t}$ expectation value under feedback control 
for finite measurement intervals (symbols) with an average of $10^2$, $10^3$, and $10^4$ trajectories (thin dotted, dashed, and solid curves in lighter colors, respectively; 
single trajectories would decay at different pace from $\pm 1$ towards 0 between measurements).
In the continuous measurement limit (top bold curve), the final purity is maximal and would vanish without feedback.
Parameters: $\lambda_+=1.0$, $\lambda_-=5.0$, $\Omega/\gamma=5.0$, $\theta_\pm=\pi/2$.
}
\end{figure}

Finally, with both dissipation and feedback, different conditioned Liouvillians are applied between the measurements.
Using the Bloch sphere representation $\rho=\frac{1}{2}\left[\f{1}+\f{r}\cdot\f{\sigma}\right]$ with $\abs{\f{r}} \le 1$ we can rephrase the fixed-point iteration for the density matrix as a 
fixed-point iteration for the expectation values of the Pauli matrices, see appendix~\ref{ASblochsphere}.
Of these, only one has a continuum limit when $\Delta t \to 0$
\bea\label{Econtsx}
\expval{\dot{\sigma}^x_t} &=& -\frac{\gamma}{4} \left\{\lambda_-^2\left[3+\cos(2\theta_-)\right]+\lambda_+^2\left[3+\cos(2\theta_+)\right]\right\} \expval{\sigma^x_t}\nn
&&+\frac{\gamma}{4} \left\{\lambda_-^2\left[3+\cos(2\theta_-)\right]-\lambda_+^2\left[3+\cos(2\theta_+)\right]\right\}\,,
\eea
whilst the other expectation values are continuously projected to zero.
First we see that under feedback, the stationary solution $\expval{\bar{\sigma}^x}$ does not vanish despite the high-temperature limit.
Remarkably, it is only weakly dependent on the interaction angles $\theta_\pm$ but is much more sensitive to the ratio 
of the dampening constants $\lambda_\pm$: The stationary state is purified when $\lambda_+\gg \lambda_-$ (where $\expval{\sigma^x_t} \to -1$)
or $\lambda_+ \ll \lambda_-$ (where $\expval{\sigma^x_t}\to+1$).
The weak dependence on the dissipation direction demonstrates that small control errors (e.g. caused by further decoherence channels)
have only small effects on purification efficiency.
When the dampening rates differ strongly (strong feedback), the eigenstate of the measurement operator that has the smaller dampening rate becomes purified.
The applicability of the effective description of measurement and feedback control by a fixed-point iteration can also be checked by comparing with averaging over many
different trajectories, see Fig.~\ref{Fsqbtraj}.

Unless either $\lambda_-$ or $\lambda_+$ vanishes (which effectively introduces a decoherence-free subspace),
the continuous measurement limit $\Delta t \to 0$ is one necessary ingredient for purification:
Then however, occurrence of purification is not specific to the chosen measurement direction.

A similiar calculation can be performed with the Hamiltonian (\ref{Ehamqubit}) for a general measurement direction 
$M_\pm = \frac{1}{2}\left[\f{1}\pm \f{n}\cdot\f{\sigma}\right]$ defined by
$\f{n}=(\sin(\theta)\cos(\phi),\sin(\theta) \sin(\phi),\cos(\theta))$, 
see appendix~\ref{ASgenmeasurement}.
Note that in this light, the Maxwell-demon type classical feedback in Ref.~\cite{schaller2011} appears as a special case where system Hamiltonian
of a single-level quantum dot $\HS=\epsilon d^\dagger d$ with fermionic operators and measurement operators for empty $M_{\rm E} = \f{1}-d^\dagger d$
and filled $M_{\rm F} = d^\dagger d$ dot states commute.
For a general measurement direction one finds for our model that the purification direction coincides with it.
The modulus of the Bloch vector of the stationary state does not depend on $\phi$ and $\phi_i$
\bea\label{Eblochvector}
\sum_{\alpha=1}^3 \expval{\bar{\sigma}^\alpha}^2 &=&
\frac{\left[\lambda_+^2 f(\theta,\theta_+)-\lambda_-^2 f(\theta,\theta_-)\right]^2}{\left[\lambda_+^2 f(\theta,\theta_+)+\lambda_-^2 f(\theta,\theta_-)\right]^2}
\eea
and only weakly on the remaining angles $f(\theta,\theta_i)=5-\cos(2\theta_i)-\cos(2\theta)\left[1+3\cos(2\theta_i)\right]$ (the special case $\theta=\theta_i=0$ corresponds to
trivial purification in the direction of $\HS$).
It is immediately evident that it approaches one (pure state) when one of the dampening coefficients $\lambda_\pm$ vanishes or is much larger than the other.

\section{Dissipative Entangling}
\begin{figure}[ht]
\includegraphics[height=0.48\textwidth,angle=-90,clip=true]{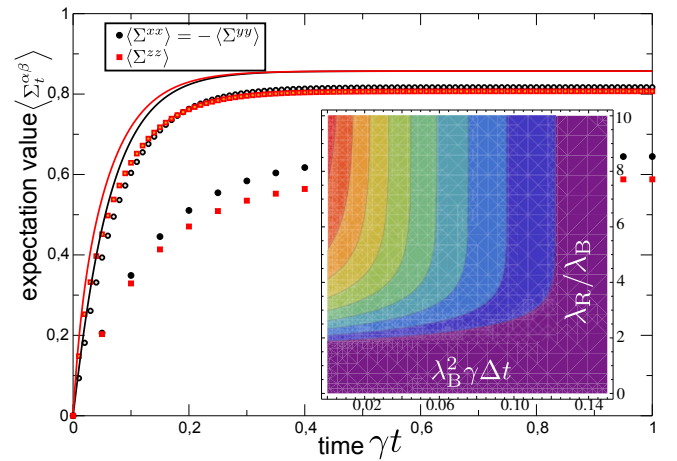}
\caption{\label{Ftwoqbtraj}(Color Online)
Effective evolution of the non-vanishing $\expval{\Sigma^{\alpha\beta}_t}$ expectation values under feedback control 
for finite measurement intervals (symbols) and in the continuum limit (solid curves).
Parameters: $\lambda_{\rm B}=1$, $\lambda_{\rm R}=5$, $\omega_A\Delta t=\omega_B\Delta t=0$.
A non-vanishing stationary concurrence (see inset extending Eq.~(\ref{Econcpur}) to finite measurement intervals, 
with contour lines ranging from 0.1 to 0.9 in steps of 0.1) requires short measurement intervals.
}
\end{figure}

For two qubits, one would be interested in a measurement scheme that leads to a purified entangled state.
A projective (non-demolition) measurement of the Bell state $\ket{\Psi_{\rm Bell}}=\left[\ket{00}+\ket{11}\right]/\sqrt{2}$
corresponds to a two-local measurement Bell (B) projection operator
\bea
M_{\rm B} = \ket{\Psi_{\rm Bell}}\bra{\Psi_{\rm Bell}}=\frac{1}{4}\left[\f{1}+\sigma^x_1\sigma^x_2-\sigma^y_1\sigma^y_2+\sigma^z_1\sigma^z_2\right]\,.\;\;
\eea
Assuming the measurement only to recognize the Bell state (two outcomes), the remaining (R) projector follows from 
the completeness relation as $M_{\rm R} = \f{1}-M_{\rm B}$.
In the energy eigenbasis defined by the local system Hamiltonian
\bea
\HS = \frac{\omega_1}{2} \sigma^z_1 + \frac{\omega_2}{2} \sigma^z_2\,,
\eea
the $16\times 16$ superoperator equivalents of $M_{\rm B}\rho M_{\rm B}^\dagger$ and $M_{\rm R} \rho M_{\rm R}^\dagger$ 
will not have a Block structure between coherences and populations, see appendix~\ref{AStwoqubit}.
For simplicity, we choose to change the strength of the interaction only
$\HI = \lambda_{\rm B/R} \left(\sigma^x_1+\sigma^x_2\right)\otimes \sum_k \left(h_k b_k+h_k^* b_k^\dagger\right)$
and to keep the bath invariant as in Eq.~(\ref{Ehamqubit}).
The resulting Liouvillians in Born-Markov-secular approximation decouple populations and coherences in the system energy eigenbasis
and simplify strongly in the high-temperature limit $\gamma_\omega=\gamma$ (where also the Lamb-shift vanishes), see appendix~\ref{AStwoqubit}.
According to Eq.~(\ref{Epropeff}), the effective propagator becomes 
${\cal P}_{\rm eff}(\Delta t) = e^{{\cal L}_{\rm B} \Delta t} {\cal M}_{\rm B} + e^{{\cal L}_{\rm R} \Delta t} {\cal M}_{\rm R}$
and defines a fixed-point iteration for the density matrix.
Just as for a single qubit one may parameterize the density matrix by Pauli matrices~\cite{bruening2012a}
\mbox{$\rho(t) = \sum_{\alpha\beta\in\{0,x,y,z\}} r^{\alpha\beta}(t) \sigma^\alpha_1 \sigma^\beta_2$}
with $\sigma^0_{1/2} \equiv \f{1}_{1/2}$ and $r^{00} = 1/4$.
The generators $\Sigma^{\alpha\beta}\equiv\sigma^\alpha_1 \sigma^\beta_2$ of the group $SU(4)$ are trace orthogonal
$\trace{\Sigma^{\alpha\beta} \Sigma^{\alpha'\beta'}} = 4 \delta_{\alpha\alpha'}\delta_{\beta\beta'}$, which allows
to obtain a fixed-point iteration for the 15 nontrivial expectation values of generalized Pauli matrices $\expval{\Sigma^{\alpha\beta}_t}=4 r^{\alpha\beta}(t)$.
The stationary state of the fixed-point iteration (also identifiable as the normalized eigenvector of the effective propagator in Eq.~(\ref{Epropeff}) with 
eigenvalue one) has in the continuum limit $\Delta t\to 0$ the only non-vanishing stationary values
\bea\label{Efixedpoint}
\expval{\bar{\Sigma}^{xx}} = - \expval{\bar{\Sigma}^{yy}} = +\expval{\bar{\Sigma}^{zz}} = \frac{\lambda_{\rm R}^2-\lambda_{\rm B}^2}{\lambda^2_{\rm R}+3\lambda^2_{\rm B}}\,,
\eea
see also appendix~\ref{ASstatstate}.
Similarly, the pure Bell state is fully characterized by the only non-vanishing expectation values $\expval{\Sigma^{xx}} = -\expval{\Sigma^{yy}} =\expval{\Sigma^{zz}}=+1$.
Eq.~(\ref{Efixedpoint}) corresponds to a stationary concurrence~\cite{wootters1998a} and purity $P=\trace{\rho^2}$ of
\bea\label{Econcpur}
\bar{C} &=& \frac{\lambda^2_{\rm R}-3\lambda^2_{\rm B}}{\lambda^2_{\rm R}+3\lambda^2_{\rm B}} \Theta(\lambda^2_{\rm R}-3\lambda^2_{\rm B})\,,\;\;
\bar{P} = \frac{\lambda_{\rm R}^4+3\lambda_{\rm B}^4}{\left(\lambda^2_{\rm R}+3\lambda^2_{\rm B}\right)^2}\qquad
\eea
with $\Theta(x)$ denoting the Heaviside step function.
Thus, when the undesired parts of the density matrix are damped much stronger than the entangled parts $\lambda_{\rm R} \gg \lambda_{\rm B}$, both
concurrence and purity approach one, which demonstrates that the method may in principle create pure maximally entangled states.
In contrast, without feedback ($\lambda_{\rm R}=\lambda_{\rm B}$), the stationary concurrence vanishes and the purity becomes that of a completely mixed two-qubit state.
For finite measurement intervals $\Delta t$, the fidelity of the state preparation reduces in comparison to Eq.~(\ref{Econcpur}), see Fig.~\ref{Ftwoqbtraj}.
A perfect stabilization of the Bell state therefore requires both the continuum limit where measurement, processing, and control are performed
much faster than decoherence,  and strong control $\lambda_{\rm R}/\lambda_{\rm B}\to\infty$.

\section{Summary}
For a piecewise-constant and instantaneous control, it is possible to explicitly include the measurement process into a quantum feedback
scheme for the evolution of an open quantum system.
However, unlike standard quantum feedback control, an effective description by a master equation is not possible anymore.
Instead, the evolution may be described by a fixed-point iteration of the density matrix.
As a very intuitive outcome, purification of eigenstates of the measurement operators can be achieved 
provided that unwanted parts of the density matrix are damped with a significantly larger rate than the desired ones.
By construction, further decoherence channels simply enter the scheme as control errors, for which we have found
only a mild effect on the purification efficiency.
The scheme is suitable for high temperatures, but generally the time interval at which measurements are performed 
has to be much smaller than the temperature-dependent decoherence time corresponding to the conditioned Lindblad evolutions.
Therefore, even when measurement-induced collapse and control are instantaneous, the purification efficiency will eventually 
be limited by the finite signal processing speed thereby for classical electronic processing requiring decoherence times in the order of milliseconds.

To achieve significant purification without a decoherence-free subspace, it is important to note that the limitation to Markovian and weakly-coupled systems
used in the derivation of the Lindblad dissipators can in the continuum measurement limit be overcome by deriving the
trace- and positivity-preserving map via coarse-graining~\cite{schaller2008a,schaller2009a},
where the coarse-graining timescale is naturally set by the detector sampling frequency $\Delta t$.
A further interesting avenue of research is how the quantum version of the Jarzinsky equality~\cite{sagawa2010a,vedral2012a} is modified by
feedback and non-projective measurements that do not commute with the system Hamiltonian.

\section{Acknowledgments}
Financial support by the DFG (SCHA 1646/2-1, SFB 910) and stimulating discussions with
T. Brandes, C. Emary, G. Kiesslich, and H. Wiseman are gratefully acknowledged.


\widetext

\appendix

\section{Jump measurement limit}\label{ASjumpmeasurement}

Suppose we consider a two-level system under the continuous action of a vaccuum reservoir and a measurement device that 
detects whether during time intervals $\Delta t$ a quantum jump has occurred in the system (e.g. decay to the ground state
via the emission of a photon).
We assume that the decay of the two-level system from its excited state $\ket{1}$ to its ground state $\ket{0}$ is detected
with unit efficiency and occurs with probability $\gamma \Delta t$.
A measurement of such quantum jumps requires a POVM description, since these are non-projective.
In addition, the measurement operators for the outcomes click (c) and no-click (nc) will depend on $\Delta t$
\bea
M_{\rm c}(\Delta t) = \sqrt{\gamma \Delta t} \ket{0}\bra{1}\,,\qquad
M_{\rm nc}(\Delta t) = \sqrt{1-\gamma\Delta t} \ket{1}\bra{1}+\ket{0}\bra{0}\,,
\eea
which automatically obeys the completeness relation $M_{\rm c}^\dagger(\Delta t) M_{\rm c}(\Delta t) + M_{\rm nc}^\dagger(\Delta t) M_{\rm nc}(\Delta t) = \f{1}$.
They act on the density matrix as
\bea
M_{\rm c}(\Delta t) \rho M_{\rm c}^\dagger(\Delta t) &=& \gamma \Delta t \rho_{11} \ket{0}\bra{0}\,,\nn
M_{\rm nc}(\Delta t) \rho M_{\rm nc}^\dagger(\Delta t) &=& \rho_{00} \ket{0}\bra{0} + (1-\gamma\Delta t)\rho_{11} \ket{1}\bra{1}+\sqrt{1-\gamma\Delta t} \rho_{01} \ket{0}\bra{1}
+\sqrt{1-\gamma\Delta t} \rho_{10} \ket{1}\bra{0}\,.
\eea
In the following, we will need their action for small $\Delta t$ and therefore state
\bea\label{Emmrel}
{\cal M}_{\rm c}(0)\rho &\hat{=}& M_{\rm c}(0) \rho M_{\rm c}^\dagger(0) = \f{0}\,,\nn 
{\cal M}_{\rm nc}(0)\rho &\hat{=}& M_{\rm nc}(0) \rho M_{\rm nc}^\dagger(0) = \rho\,,\nn
{\cal M}_{\rm c}'(0)\rho &\hat{=}& \frac{d}{d\Delta t} \left[M_{\rm c}(\Delta t) \rho M_{\rm c}^\dagger(\Delta t)\right]_{\Delta t=0} = \gamma \rho_{11} \ket{0}\bra{0}\,,\nn
{\cal M}_{\rm nc}'(0)\rho &\hat{=}& \frac{d}{d\Delta t} \left[M_{\rm c}(\Delta t) \rho M_{\rm c}^\dagger(\Delta t)\right]_{\Delta t=0} =
-\gamma \rho_{11} \ket{1}\bra{1} - \frac{\gamma}{2} \rho_{01} \ket{0}\bra{1} - \frac{\gamma}{2} \rho_{10} \ket{1} \bra{0}\,.
\eea
The feedback scheme can now be defined by performing an instantaneous unitary transformation $U_{\rm c}$ 
(experimentally approximated by using a $\delta$-kick like pulse on the system Hamiltonian)
of the density matrix whenever a detector click has occured.
In addition, the system Hamiltonian (and possible further unmonitored reservoirs) is included in the Lindblad superoperator ${\cal L}_0$.
Then, the effective propagator becomes
\bea
{\cal P}(\Delta t) = e^{{\cal L}_0 \Delta t} e^{\kappa_{\rm c}} {\cal M}_{\rm c}(\Delta t) + e^{{\cal L}_0 \Delta t} {\cal M}_{\rm nc}(\Delta t)\,,
\eea
where $e^{\kappa_{\rm c}}\rho \equiv U_{\rm c} \rho U_{\rm c}^\dagger$ and correspondence to Eq.~(\ref{Epropeffgeneral}) is established by 
putting ${\cal B}_{\rm c}(\Delta t) = e^{{\cal L}_0\Delta t} e^{\kappa_{\rm c}}$
and ${\cal B}_{\rm nc}(\Delta t) = e^{{\cal L}_0\Delta t}$.
For small $\Delta t$ (continuum limit), this allows to define an effective master equation under jump detection and unitary control
\bea
\dot{\rho} &=& \lim\limits_{\Delta t\to 0} \frac{\rho(t+\Delta t)-\rho(t)}{\Delta t} 
= \lim\limits_{\Delta t\to 0} \frac{1}{\Delta t} \left[{\cal P}(\Delta t) - \f{1}\right]\rho(t)\nn
&=&  \lim\limits_{\Delta t\to 0} \frac{1}{\Delta t} \left\{
\left[e^{\kappa_{\rm c}} {\cal M}_{\rm c}(0) + {\cal M}_{\rm nc}(0) -\f{1}\right]
+ \Delta t\left[{\cal L}_0 e^{\kappa_{\rm c}} {\cal M}_{\rm c}(0) + {\cal L}_0 {\cal M}_{\rm nc}(0)
+ e^{\kappa_{\rm c}} {\cal M}_{\rm c}'(0) + {\cal M}_{\rm nc}'(0)\right]\right\}\rho(t)\nn
&=& \left[+{\cal L}_0 + e^{\kappa_{\rm c}} {\cal M}_{\rm c}'(0) + {\cal M}_{\rm nc}'(0)\right] \rho(t) = {\cal L}_{\rm fb} \rho(t) \hat{=} {\cal L}_{\rm fb}[\rho(t)]\,,
\eea
where we have already used the superoperator equivalents of Eq.~(\ref{Emmrel}).
At the second line, a necessary ingredient for an effective master equation description becomes obvious: To remove the first term in 
square brackets, it is formally required that $\sum_m {\cal B}_m(0) {\cal M}_m(0) = \f{1}$.
Here, this is fulfilled since infinitely short measurements do not affect the density matrix.
Finally, also inserting the derivatives of the measurement superoperators makes the Liouvillian in the effective master equation picture explicit
\bea
{\cal L}_{\rm fb}[\rho] &=&  {\cal L}_0[\rho] + \gamma \left[U_{\rm c} \ket{0}\bra{1}\rho \ket{1}\bra{0}U_{\rm c}^\dagger - \frac{1}{2}\left\{\ket{1}\bra{1}, \rho\right\}\right]\nn
&=& {\cal L}_0[\rho] + \gamma \left[L_{\rm c} \rho L_{\rm c}^\dagger - \frac{1}{2} \left\{L_{\rm c}^\dagger L_{\rm c}, \rho\right\}\right]
\eea
with $L_{\rm c} = U_{\rm c} \ket{0}\bra{1}$, which obviously is a master equation of Lindblad type.


\section{Single-qubit Liouvillian}\label{ASsinglequbit}

The Born-, Markov-, and secular approximations lead for a non-degenerate system Hamiltonian generally 
to a decoupling of the evolution equations for populations and coherences in the system energy eigenbasis.
For the single qubit with $\Omega>0$, non-degeneracy is obviously fulfilled.
Starting from Eq.~(\ref{Ehamqubit}) in the main text, the Liouvillians act on the density matrix elements $\rho_{ij} = \bra{i}\rho\ket{j}$
as
\bea
\dot{\rho}_{00} &=& -\lambda_\pm^2 \sin^2(\theta_\pm) \gamma_{+\Omega} \rho_{00} + \lambda_\pm^2 \sin^2(\theta_\pm) \gamma_{-\Omega} \rho_{11}\,,\nn
\dot{\rho}_{11} &=& +\lambda_\pm^2 \sin^2(\theta_\pm) \gamma_{+\Omega} \rho_{00} - \lambda_\pm^2 \sin^2(\theta_\pm) \gamma_{-\Omega} \rho_{11}\,,\nn
\dot{\rho}_{01} &=& \left[-\ii \Omega -\lambda_\pm^2 \sin^2(\theta_\pm) \left(\sigma_{+\Omega}-\sigma_{-\Omega}\right)/2 
-\lambda_\pm^2 \sin^2(\theta_\pm) \left(\gamma_{+\Omega}-\gamma_{-\Omega}\right)/2 - 2 \lambda_\pm^2 \cos^2(\theta_\pm) \gamma_0\right] \rho_{01}\,,\nn
\dot{\rho}_{10} &=& \left[+\ii \Omega +\lambda_\pm^2 \sin^2(\theta_\pm) \left(\sigma_{+\Omega}-\sigma_{-\Omega}\right)/2 
-\lambda_\pm^2 \sin^2(\theta_\pm) \left(\gamma_{+\Omega}-\gamma_{-\Omega}\right)/2 - 2 \lambda_\pm^2 \cos^2(\theta_\pm) \gamma_0\right] \rho_{10}\,,
\eea
where $\gamma_\omega$ is the even (real-valued) Fourier transform of the bath correlation function
\bea
C(\tau) &=& \expval{e^{+\ii \HB \tau} B e^{-\ii \HB \tau} B}
= \sum_{kk'}\expval{ \left[h_k b_k e^{-\ii\omega_k \tau} + h_k^* b_k^\dagger e^{+\ii\omega_k \tau}\right]\left[h_{k'} b_{k'} + h_{k'}^* b_{k'}^\dagger\right]}\nn
&=& \sum_k \abs{h_k}^2 \left[e^{-\ii\omega_k \tau} \left[1+n_{\rm B}(\omega_k)\right] + e^{+\ii\omega_k\tau} n_{\rm B}(\omega_k)\right]
= \frac{1}{2\pi} \int\limits_0^\infty J(\omega) \left[e^{-\ii\omega\tau} \left[1+n_{\rm B}(\omega)\right] + e^{+\ii\omega\tau} n_{\rm B}(\omega)\right] d\omega\nn
&=&  \frac{1}{2\pi} \int\limits_{-\infty}^{+\infty} J(\omega)  \left[1+n_{\rm B}(\omega)\right] e^{-\ii\omega\tau} d\omega
\eea
as defined in the main text and similarly the imaginary quantity
\bea
\sigma_\omega = \int \expval{e^{+\ii \HB \tau} B e^{-\ii \HB \tau} B} e^{+\ii \omega\tau} \sgn{\tau} d\tau = \frac{\ii}{\pi} {\cal P} \int \frac{\gamma_{\bar{\omega}}}{\omega-\bar{\omega}} d\bar{\omega}
\eea
denotes its odd Fourier transform, which may also be obtained from the even one via a Cauchy principal value integral.
The odd Fourier transform is associated with the Lamb-shift terms that account for an energy renormalization of the qubit.
Note that in the high-temperature and wide-band limits that we focus on, $\gamma_\omega$ is approximately flat, such that $\sigma_\omega \approx 0$.
Then, the stationary state of both Liouvillians is the completely mixed one with $\bar{\rho}_{00} = \bar{\rho}_{11} = 1/2$ and $\bar{\rho}_{01}=\bar{\rho}_{10}=0$.
However, $\sigma_\omega$ is purely imaginary in any case, which for the evolution of the absolute square of coherences implies Eq.~(\ref{Edephrate}) in the main text.
Similarly, we obtain thermalization from the evolution of populations by invoking the KMS condition.


\section{Bloch sphere iteration}\label{ASblochsphere}

The Bloch sphere representation of the density matrix $\rho = \frac{1}{2} \left[\f{1} + \f{r} \cdot \f{\sigma}\right]$ implies $\expval{\sigma^\alpha} = r^\alpha$ or
alternatively for the density vector the representation
\bea
\rho = \left(
\frac{1}{2}\left(1+\expval{\sigma^z}\right), \frac{1}{2}\left(1-\expval{\sigma^z}\right), \frac{1}{2}\left(\expval{\sigma^x}-\ii\expval{\sigma^y}\right), \frac{1}{2}\left(\expval{\sigma^x}+\ii\expval{\sigma^y}\right)
\right)^{\rm T}\,,
\eea
which can be inserted in the fixed-point iteration for the density matrix $\rho(t+\Delta t) = {\cal P}_{\rm eff}(\Delta t) \rho(t)$ to yield iteration equations for the expectation values 
of Pauli matrices
\bea
\expval{\sigma^x_{t+\Delta t}} &=& \frac{\cos(\Omega\Delta t)}{2} \left(e^{-\gamma\Delta t\lambda_+^2[3+\cos(2\theta_+)]/2}-e^{-\gamma\Delta t\lambda_-^2[3+\cos(2\theta_-)]/2}\right)\nn
&&+ \frac{\cos(\Omega\Delta t)}{2} \left(e^{-\gamma\Delta t\lambda_+^2[3+\cos(2\theta_+)]/2}+e^{-\gamma\Delta t\lambda_-^2[3+\cos(2\theta_-)]/2}\right) \expval{\sigma^x_t}\,,\nn
\expval{\sigma^y_{t+\Delta t}} &=& \frac{\sin(\Omega\Delta t)}{2} \left(e^{-\gamma\Delta t\lambda_+^2[3+\cos(2\theta_+)]/2}-e^{-\gamma\Delta t\lambda_-^2[3+\cos(2\theta_-)]/2}\right)\nn
&&+ \frac{\sin(\Omega\Delta t)}{2} \left(e^{-\gamma\Delta t\lambda_+^2[3+\cos(2\theta_+)]/2}+e^{-\gamma\Delta t\lambda_-^2[3+\cos(2\theta_-)]/2}\right) \expval{\sigma^x_t}\,,\nn
\expval{\sigma^z_{t+\Delta t}} &=& 0\,.
\eea
The $\expval{\sigma^z}$-expectation values does not recover from zero after the first projection since the chosen system Hamiltonian was proportional to $\sigma^z$, which only generates
rotations between $\sigma^x$ and $\sigma^y$.
These equations recover the no-feedback limit ($\lambda_\pm = \lambda$ and $\theta_\pm = \theta$) disscussed and the continuum limit in Eq.~(\ref{Econtsx}) in the main text.


\section{Most general measurement direction}\label{ASgenmeasurement}

For more general projective measurements $M_\pm = \frac{1}{2}\left[\f{1}\pm \f{n} \cdot \f{\sigma}\right]=M_\pm^\dagger$, their action on the density
vector in ordering $\rho=(\rho_{00},\rho_{11},\rho_{01},\rho_{10})^{\rm T}$ is given by
\bea
{\cal M}_+ &=& \frac{1}{4}\left(\begin{array}{cccc}
4 \cos^4\left(\frac{\theta}{2}\right) & \sin^2(\theta) & 
+4 e^{+\ii\phi} \sin\left(\frac{\theta}{2}\right) \cos^3\left(\frac{\theta}{2}\right) &  +4 e^{-\ii\phi} \sin\left(\frac{\theta}{2}\right) \cos^3\left(\frac{\theta}{2}\right)\\
\sin^2(\theta) & 4 \sin^4\left(\frac{\theta}{2}\right) & 
+4 e^{+\ii\phi} \sin^3\left(\frac{\theta}{2}\right) \cos\left(\frac{\theta}{2}\right) &  +4 e^{-\ii\phi} \sin^3\left(\frac{\theta}{2}\right) \cos\left(\frac{\theta}{2}\right)\\
+4 e^{-\ii\phi} \sin\left(\frac{\theta}{2}\right) \cos^3\left(\frac{\theta}{2}\right) & +4 e^{-\ii\phi} \sin^3\left(\frac{\theta}{2}\right) \cos\left(\frac{\theta}{2}\right) & 
\sin^2(\theta) & e^{-2\ii\phi} \sin^2(\theta)\\
+4 e^{+\ii\phi} \sin\left(\frac{\theta}{2}\right) \cos^3\left(\frac{\theta}{2}\right) & +4 e^{+\ii\phi} \sin^3\left(\frac{\theta}{2}\right) \cos\left(\frac{\theta}{2}\right) & 
e^{+2\ii\phi} \sin^2(\theta) & \sin^2(\theta)
\end{array}\right)\,,\nn
{\cal M}_- &=& \frac{1}{4}\left(\begin{array}{cccc}
4 \sin^4\left(\frac{\theta}{2}\right) & \sin^2(\theta) & 
-4 e^{+\ii\phi} \sin^3\left(\frac{\theta}{2}\right) \cos\left(\frac{\theta}{2}\right) & -4 e^{-\ii\phi} \sin^3\left(\frac{\theta}{2}\right) \cos\left(\frac{\theta}{2}\right)\\
\sin^2(\theta) & 4 \cos^4\left(\frac{\theta}{2}\right) & 
-4 e^{+\ii\phi} \sin\left(\frac{\theta}{2}\right) \cos^3\left(\frac{\theta}{2}\right) & -4 e^{-\ii\phi} \sin\left(\frac{\theta}{2}\right) \cos^3\left(\frac{\theta}{2}\right)\\
-4 e^{-\ii\phi} \sin^3\left(\frac{\theta}{2}\right) \cos\left(\frac{\theta}{2}\right) & -4 e^{-\ii\phi} \sin\left(\frac{\theta}{2}\right) \cos^3\left(\frac{\theta}{2}\right) & 
\sin^2(\theta) & e^{-2\ii\phi} \sin^2(\theta)\\
-4 e^{+\ii\phi} \sin^3\left(\frac{\theta}{2}\right) \cos\left(\frac{\theta}{2}\right) & -4 e^{+\ii\phi} \sin\left(\frac{\theta}{2}\right) \cos^3\left(\frac{\theta}{2}\right) & 
e^{+2\ii\phi} \sin^2(\theta) & \sin^2(\theta)
\end{array}\right)\,.
\eea
One can easily check the orthogonal projector properties ${\cal M}_\pm^2 = {\cal M}_\pm$ and ${\cal M}_\pm {\cal M}_\mp = \f{0}$.
In addition, the special case of Eq.~(4) in the main text is obtained by putting $\theta=\pi/2$ and $\phi=0$.
Inserting the corresponding effective propagator into the fixed-point iteration for the Pauli matrix expectation values one now observes that all expectation values couple to each other.
Solving for the stationary state yields Eq.~(\ref{Eblochvector}) of the main text.
Naturally, Eq.~(\ref{Eblochvector}) is also compatible with the stationary state of Eq.~(\ref{Econtsx}) when $\theta=\pi/2$ and $\phi=0$.

\section{Two-qubit Liouvillian}\label{AStwoqubit}

In the energy eigenbasis of the two qubits $\ket{00}$, $\ket{01}$, $\ket{10}$, $\ket{11}$, the Born-, Markov-, and secular approximations again
lead to a decoupled evolution of populations and coherences. The populations ($\rho_{ij,k\ell} = \bra{ij}\rho\ket{k\ell}$) evolve
according to
\bea
\dot{\rho}_{00,00} &=& -(\lambda_{\rm B/R}^2 \gamma_{+\omega_1}+\lambda_{\rm B/R}^2 \gamma_{+\omega_2})\rho_{00,00}+\lambda_{\rm B/R}^2 \gamma_{-\omega_2} \rho_{01,01} +\lambda_{\rm B/R}^2 \gamma_{-\omega_1} \rho_{10,10}\,,\nn
\dot{\rho}_{01,01} &=& +\lambda_{\rm B/R}^2 \gamma_{+\omega_2} \rho_{00,00} -(\lambda_{\rm B/R}^2 \gamma_{+\omega_1}+\lambda_{\rm B/R}^2 \gamma_{-\omega_2}) \rho_{01,01} + \lambda_{\rm B/R}^2 \gamma_{-\omega_1} \rho_{11,11}\,,\nn
\dot{\rho}_{10,10} &=& +\lambda_{\rm B/R}^2 \gamma_{+\omega_1} \rho_{00,00} -(\lambda_{\rm B/R}^2 \gamma_{-\omega_1}+\lambda_{\rm B/R}^2 \gamma_{+\omega_2}) \rho_{10,10} + \lambda_{\rm B/R}^2 \gamma_{-\omega_2} \rho_{11,11}\,,\nn
\dot{\rho}_{11,11} &=& +\lambda_{\rm B/R}^2 \gamma_{+\omega_1} \rho_{01,01} +\lambda_{\rm B/R}^2 \gamma_{+\omega_2} \rho_{10,10} -(\lambda_{\rm B/R}^2 \gamma_{-\omega_1}+\lambda_{\rm B/R}^2 \gamma_{-\omega_2}) \rho_{11,11}
\eea
and the coherences decay independently (omitted for brevity).
By invoking the KMS condition $\gamma_{-\omega} = e^{-\beta\omega} \gamma_{+\omega}$ it is immediately obvious that the stationary state is the thermalized one.
In the high-temperature limit considered in the paper, the stationary state of either stationary Liouvillian is just the completely mixed state.
In this limit, the coherences decay according to
\bea
\dot{\rho}_{00,01} &=& \left(-\ii\omega_2-2\lambda_{\rm B/R}^2\gamma\right)\rho_{00,01}+\lambda_{\rm B/R}^2\gamma\rho_{10,11}\,,\nn
\dot{\rho}_{00,10} &=& \left(-\ii\omega_1-2\lambda_{\rm B/R}^2\gamma\right)\rho_{00,10}+\lambda_{\rm B/R}^2\gamma\rho_{01,11}\,,\nn
\dot{\rho}_{00,11} &=& \left(-\ii\omega_1-\ii\omega_2-2\lambda_{\rm B/R}^2\gamma\right)\rho_{00,11}\,,\nn
\dot{\rho}_{01,00} &=& \left(+\ii\omega_2-2\lambda_{\rm B/R}^2\gamma\right)\rho_{01,00}+\lambda_{\rm B/R}^2\gamma\rho_{11,10}\,,\nn
\dot{\rho}_{01,10} &=& \left(-\ii\omega_1+\ii\omega_2-2\lambda_{\rm B/R}^2\gamma\right)\rho_{01,10}\,,\nn
\dot{\rho}_{01,11} &=& \left(-\ii\omega_1-2\lambda_{\rm B/R}^2\gamma\right)\rho_{01,11}+\lambda_{\rm B/R}^2\gamma\rho_{00,10}\,,\nn
\dot{\rho}_{10,00} &=& \left(+\ii\omega_1-2\lambda_{\rm B/R}^2\gamma\right)\rho_{10,00}+\lambda_{\rm B/R}^2\gamma\rho_{11,01}\,,\nn
\dot{\rho}_{10,01} &=& \left(+\ii\omega_1-\ii\omega_2-2\lambda_{\rm B/R}^2\gamma\right)\rho_{10,01}\,,\nn
\dot{\rho}_{10,11} &=& \left(-\ii\omega_2-2\lambda_{\rm B/R}^2\gamma\right)\rho_{10,11}+\lambda_{\rm B/R}^2\gamma\rho_{00,01}\,,\nn
\dot{\rho}_{11,00} &=& \left(+\ii\omega_1+\ii\omega_2-2\lambda_{\rm B/R}^2\gamma\right)\rho_{11,00}\,,\nn
\dot{\rho}_{11,01} &=& \left(+\ii\omega_1-2\lambda_{\rm B/R}^2\gamma\right)\rho_{11,01}+\lambda_{\rm B/R}^2\gamma\rho_{10,00}\,,\nn
\dot{\rho}_{11,10} &=& \left(+\ii\omega_2-2\lambda_{\rm B/R}^2\gamma\right)\rho_{11,10}+\lambda_{\rm B/R}^2\gamma\rho_{01,00}\,,
\eea
where coherences may couple among themselves provided they belong to superpositions of states with equal energy differences.
Inserting $\gamma_\omega\to\gamma$ also in the equations for the populations yields together with the sparse $16\times 16$ measurement superoperators
\bea
{\cal M}_B &=&
\frac{1}{4}\left(\begin{array}{cccccccccccccccc}
1 & 0 & 0 & 1 & 0 & 0 & 1 & 0 & 0 & 0 & 0 & 0 & 0 & 1 & 0 & 0\\
0 & 0 & 0 & 0 & 0 & 0 & 0 & 0 & 0 & 0 & 0 & 0 & 0 & 0 & 0 & 0\\
0 & 0 & 0 & 0 & 0 & 0 & 0 & 0 & 0 & 0 & 0 & 0 & 0 & 0 & 0 & 0\\
1 & 0 & 0 & 1 & 0 & 0 & 1 & 0 & 0 & 0 & 0 & 0 & 0 & 1 & 0 & 0\\
0 & 0 & 0 & 0 & 0 & 0 & 0 & 0 & 0 & 0 & 0 & 0 & 0 & 0 & 0 & 0\\
0 & 0 & 0 & 0 & 0 & 0 & 0 & 0 & 0 & 0 & 0 & 0 & 0 & 0 & 0 & 0\\
1 & 0 & 0 & 1 & 0 & 0 & 1 & 0 & 0 & 0 & 0 & 0 & 0 & 1 & 0 & 0\\
0 & 0 & 0 & 0 & 0 & 0 & 0 & 0 & 0 & 0 & 0 & 0 & 0 & 0 & 0 & 0\\
0 & 0 & 0 & 0 & 0 & 0 & 0 & 0 & 0 & 0 & 0 & 0 & 0 & 0 & 0 & 0\\
0 & 0 & 0 & 0 & 0 & 0 & 0 & 0 & 0 & 0 & 0 & 0 & 0 & 0 & 0 & 0\\
0 & 0 & 0 & 0 & 0 & 0 & 0 & 0 & 0 & 0 & 0 & 0 & 0 & 0 & 0 & 0\\
0 & 0 & 0 & 0 & 0 & 0 & 0 & 0 & 0 & 0 & 0 & 0 & 0 & 0 & 0 & 0\\
0 & 0 & 0 & 0 & 0 & 0 & 0 & 0 & 0 & 0 & 0 & 0 & 0 & 0 & 0 & 0\\
1 & 0 & 0 & 1 & 0 & 0 & 1 & 0 & 0 & 0 & 0 & 0 & 0 & 1 & 0 & 0\\
0 & 0 & 0 & 0 & 0 & 0 & 0 & 0 & 0 & 0 & 0 & 0 & 0 & 0 & 0 & 0\\
0 & 0 & 0 & 0 & 0 & 0 & 0 & 0 & 0 & 0 & 0 & 0 & 0 & 0 & 0 & 0
\end{array}\right)
\eea
and 
\bea
{\cal M}_R &=&
\frac{1}{4}\left(\begin{array}{cccccccccccccccc}
1 & 0 & 0 & 1 & 0 & 0 & -1 & 0 & 0 & 0 & 0 & 0 & 0 & -1 & 0 & 0\\
0 & 4 & 0 & 0 & 0 & 0 & 0 & 0 & 0 & 0 & 0 & 0 & 0 & 0 & 0 & 0\\
0 & 0 & 4 & 0 & 0 & 0 & 0 & 0 & 0 & 0 & 0 & 0 & 0 & 0 & 0 & 0\\
1 & 0 & 0 & 1 & 0 & 0 & -1 & 0 & 0 & 0 & 0 & 0 & 0 & -1 & 0 & 0\\
0 & 0 & 0 & 0 & 2 & 0 & 0 & 0 & 0 & 0 & 0 & 0 & 0 & 0 & -2 & 0\\
0 & 0 & 0 & 0 & 0 & 2 & 0 & 0 & 0 & 0 & 0 & 0 & 0 & 0 & 0 & -2\\
-1 & 0 & 0 & -1 & 0 & 0 & 1 & 0 & 0 & 0 & 0 & 0 & 0 & 1 & 0 & 0\\
0 & 0 & 0 & 0 & 0 & 0 & 0 & 2 & 0 & -2 & 0 & 0 & 0 & 0 & 0 & 0\\
0 & 0 & 0 & 0 & 0 & 0 & 0 & 0 & 4 & 0 & 0 & 0 & 0 & 0 & 0 & 0\\
0 & 0 & 0 & 0 & 0 & 0 & 0 & -2 & 0 & 2 & 0 & 0 & 0 & 0 & 0 & 0\\
0 & 0 & 0 & 0 & 0 & 0 & 0 & 0 & 0 & 0 & 2 & 0 & -2 & 0 & 0 & 0\\
0 & 0 & 0 & 0 & 0 & 0 & 0 & 0 & 0 & 0 & 0 & 4 & 0 & 0 & 0 & 0\\
0 & 0 & 0 & 0 & 0 & 0 & 0 & 0 & 0 & 0 & -2 & 0 & 2 & 0 & 0 & 0\\
-1 & 0 & 0 & -1 & 0 & 0 & 1 & 0 & 0 & 0 & 0 & 0 & 0 & 1 & 0 & 0\\
0 & 0 & 0 & 0 & -2 & 0 & 0 & 0 & 0 & 0 & 0 & 0 & 0 & 0 & 2 & 0\\
0 & 0 & 0 & 0 & 0 & -2 & 0 & 0 & 0 & 0 & 0 & 0 & 0 & 0 & 0 & 2
\end{array}\right)
\eea
acting from the left on density vectors arranged as\\
$\Big(\rho_{00,00},\rho_{01,01},\rho_{10,10},\rho_{11,11},\rho_{00,01},\rho_{00,10},\rho_{00,11},\rho_{01,00},\rho_{01,10},\rho_{01,11},
\rho_{10,00},\rho_{10,01},\rho_{10,11},\rho_{11,00},\rho_{11,01},\rho_{11,10}\Big)^{\rm T}$,
the effective propagator for the fixed-point iteration of the density matrix.


\section{Stationary state of two-qubit fixed-point iteration}\label{ASstatstate}

Mapping the fixed-point iteration for the density matrix to the expectation values of $\expval{\Sigma^{\alpha\beta}}$, one obtains
lengthy expressions for the 15 nontrivial expectation values:
$\expval{\sigma^x_2}$, $\expval{\sigma^y_2}$, $\expval{\sigma^z_2}$, $\expval{\sigma^x_1}$, $\expval{\sigma^y_1}$, $\expval{\sigma^z_1}$, 
$\expval{\sigma^x_1 \sigma^x_2}$, $\expval{\sigma^x_1 \sigma^y_2}$, $\expval{\sigma^x_1 \sigma^z_2}$, 
$\expval{\sigma^y_1 \sigma^x_2}$, $\expval{\sigma^y_1 \sigma^y_2}$, $\expval{\sigma^y_1 \sigma^z_2}$, 
$\expval{\sigma^z_1 \sigma^x_2}$, $\expval{\sigma^z_1 \sigma^y_2}$, and $\expval{\sigma^z_1 \sigma^z_2}$. 
These can be solved for the stationary state.
Alternatively, the stationary state of the iteration can be obtained directly by 
looking at the trace-normalized eigenvector of the effective propagator with eigenvalue one.
For finite $\Delta t$, it is characterized 
by vanishing expectation values of all local operators.
The non-vanishing stationary expectation values read
\bea
\expval{\sigma^x_1 \sigma^y_2} &=& 
\frac{4 e^{3(\Lambda_{\rm B}+\Lambda_{\rm R})}\sinh\left[\Lambda_{\rm B}-\Lambda_{\rm R}\right]\sinh\left[2\Lambda_{\rm R}\right] \sin(\Omega)}
{e^{6\Lambda_{\rm R}}+e^{4\Lambda_{\rm B}+2\Lambda_{\rm R}}\left(3-4e^{4\Lambda_{\rm R}}\right)+\left(e^{2\Lambda_{\rm B}}
+e^{2\Lambda_{\rm R}}\right)\left(2e^{2\Lambda_{\rm B}+4\Lambda_{\rm R}}-e^{2\Lambda_{\rm B}}-e^{2\Lambda_{\rm R}}\right)\cos(\Omega)}\,,\nn
\expval{\sigma^x_1 \sigma^x_2} &=&\frac{4 e^{3(\Lambda_{\rm B}+\Lambda_{\rm R})}\sinh\left[\Lambda_{\rm B}-\Lambda_{\rm R}\right]\sinh\left[2\Lambda_{\rm R}\right] \cos(\Omega)}
{e^{6\Lambda_{\rm R}}+e^{4\Lambda_{\rm B}+2\Lambda_{\rm R}}\left(3-4e^{4\Lambda_{\rm R}}\right)+\left(e^{2\Lambda_{\rm B}}
+e^{2\Lambda_{\rm R}}\right)\left(2e^{2\Lambda_{\rm B}+4\Lambda_{\rm R}}-e^{2\Lambda_{\rm B}}-e^{2\Lambda_{\rm R}}\right)\cos(\Omega)}\,,\nn
\expval{\sigma^z_1 \sigma^z_2} &=&
\frac{\left(e^{4\Lambda_{\rm R}}-e^{4\Lambda_{\rm B}}\right)\left(e^{2\Lambda_{\rm R}}-\cos(\Omega)\right)}
{-e^{6\Lambda_{\rm R}}+e^{4\Lambda_{\rm B}+2\Lambda_{\rm R}}\left(4e^{4\Lambda_{\rm R}}-3\right)+\left[e^{4\Lambda_{\rm B}}+e^{4\Lambda_{\rm R}}-2e^{4\Lambda_{\rm B}+4\Lambda_{\rm R}}
-2e^{2\Lambda_{\rm B}+2\Lambda_{\rm R}}\left(e^{4\Lambda_{\rm R}}-1\right)\right]\cos(\Omega)}\,,\nn
\expval{\sigma^y_1 \sigma^x_2} &=& \expval{\sigma^x_1 \sigma^y_2}\,,\qquad
\expval{\sigma^y_1 \sigma^y_2} = -\expval{\sigma^x_1 \sigma^x_2}\,,
\eea
where we have used the dimensionless quantities $\Omega\equiv(\omega_1+\omega_2)\Delta t$, $\Lambda_{\rm B}\equiv\gamma\Delta t \lambda_{\rm B}^2$, 
and $\Lambda_{\rm R}\equiv\gamma\Delta t \lambda_{\rm R}^2$.
In the limit $\Delta t\to 0$, this recovers Eq.~(11) in the main text.
Even for finite $\Delta t$ however, the stationary density matrix is just given by three independent parameters
$\alpha_1=\expval{\sigma^x_1\sigma^y_2}=\expval{\sigma^y_1\sigma^x_2}$, $\alpha_2=\expval{\sigma^x_1\sigma^x_2}=-\expval{\sigma^y_1\sigma^y_2}$, and $\alpha_3=\expval{\sigma^z_1\sigma^z_2}$, for 
which one may also express concurrence and purity analytically which generalizes Eq.~(\ref{Econcpur}) in the main text (not shown).


\end{document}